\documentclass[12pt]{article}

\usepackage{epsfig}
\usepackage{times}
\renewcommand{\epsilon}{\varepsilon}

\newenvironment{sciabstract}{%
\begin{quote} \bf}
{\end{quote}}

\newcounter{lastnote}
\newenvironment{scilastnote}{%
\setcounter{lastnote}{\value{enumiv}}%
\addtocounter{lastnote}{+1}%
\begin{list}%
{\arabic{lastnote}.} {\setlength{\leftmargin}{.22in}}
{\setlength{\labelsep}{.5em}}} {\end{list}}

\title{Mach-Zehnder Interferometry in a Strongly Driven Superconducting Qubit}

\author
 {William D. Oliver,$^{1\ast}$ Yang Yu,$^{2}$ Janice C. Lee,$^{2}$ Karl K. Berggren,$^{2}$ \\
    Leonid S. Levitov,$^{3}$ Terry P. Orlando$^{2}$\\
\\
 \normalsize{$^{1}$MIT Lincoln Laboratory, 244 Wood Street, Lexington, MA 02420, USA}\\
 \normalsize{$^{2}$Department of Electrical Engineering and Computer Science,} \\
 \normalsize{Massachusetts Institute of Technology, Cambridge MA 02139, USA}\\
 \normalsize{$^{3}$Department of Physics, Massachusetts Institute of Technology, Cambridge MA 02139, USA}\\
 \\
 \normalsize{$^\ast$To whom correspondence should be addressed; E-mail:  oliver@ll.mit.edu.} }

\date{}

\begin{document}


\baselineskip24pt


\maketitle


\begin{sciabstract}
We demonstrate Mach-Zehnder-type interferometry in a
superconducting flux qubit. The qubit is a tunable artificial
atom, whose ground and excited states exhibit an avoided crossing.
Strongly driving the qubit with harmonic excitation sweeps it
through the avoided crossing two times per period. As the induced
Landau-Zener transitions act as coherent beamsplitters, the
accumulated phase between transitions, which varies with microwave
amplitude, results in quantum interference fringes for $n=1 \ldots
20$ photon transitions. The generalization of optical Mach-Zehnder
interferometry, performed in qubit phase space, provides an
alternative means to manipulate and characterize the qubit in the
strongly-driven regime.
\end{sciabstract}

\newpage

The development of artificial atoms with lithographically defined
superconducting circuits presents a new paradigm of quantum solid
state physics~\cite{Makhlin01a}, allowing the realization and
exploration of new macroscopic quantum
phenomena~\cite{Nakamura99a,Friedman00a,Wal00a,Nakamura01,Vion02a,Yu02a,Martinis02a,Chiorescu03a},
and also holding promise for applications in quantum
computing~\cite{NielsenChuang}. Of the various effects
demonstrated with qubits, of greatest importance are
time-dependent coherent phenomena. Those include the observation
of Rabi oscillations in charge, flux, and phase
qubits~\cite{Nakamura99a,Nakamura01,Vion02a,Yu02a,Martinis02a,Chiorescu03a},
entanglement of two~\cite{Berkley03a} and three~\cite{Xu05a}
qubits, coherent oscillation~\cite{Claudon04a} and
bifurcation~\cite{Siddiqi04} in multilevel systems, and the
demonstration of basic elements of coherent
control~\cite{Pashkin03a,Yamamoto03a,Mcdermott05a}. Artificial
atoms strongly coupled to photons have opened the arena of
``circuit quantum electrodynamics'' (c-QED)
\cite{Chiorescu04,Wallraff04}.

Here we demonstrate an application of superconducting qubits to
quantum physics, realized in a strongly driven flux qubit, and
described in terms of a Mach-Zehnder (MZ) interferometer. The
conventional MZ setup uses two beamsplitters: the first divides an
optical signal into two coherent waves which travel along paths
with different effective lengths, and the second recombines and
superposes these waves, leading to quantum interference fringes in
the measured output signal. In a driven qubit, following an idea
discussed in Ref.~\cite{Shytov03}, the beamsplitters can be
realized by Landau-Zener (LZ) transitions at a level avoided
crossing. Over one oscillation period of the driving field, the
qubit is swept through the avoided crossing twice
(Fig.~\ref{fig:MZinterferometer}A). Starting from the marker, at
the first LZ transition (time $t_1$), the ground state $|0\rangle$
is split into a coherent superposition of the ground and excited
states, $|1\rangle$ and $|0\rangle$, which, after evolving
independently and accumulating a relative phase
$\Delta\theta_{12}$, interfere at the second LZ transition (time
$t_2$). The corresponding qubit-state energy evolution (first
period, Fig.~\ref{fig:MZinterferometer}B) between the recurrent LZ
transitions (shaded region) provides a phase-space analog to the
two arms and the beamsplitters of an optical MZ interferometer
(top left, Fig,~\ref{fig:MZinterferometer}B). The interference
phase
\begin{equation}\label{eq:phase}
 \Delta\theta_{12}=\frac1{\hbar}\int_{t_1}^{t_2} \epsilon(t)dt ,\quad \epsilon(t)=\epsilon_{|0\rangle}(t)-\epsilon_{|1\rangle}(t) ,
\end{equation}
depends on the magnitude of the qubit energy detuning excursion
for times $t_1<t<t_2$. The interference fringes in the occupation
probability correspond to integer and half-integer values of
$\Delta\theta_{12}/2\pi$. Known as St\"uckelberg oscillations with
Rydberg atoms~\cite{Baruch92,Yoakum92}, this mechanism has
application to quantum control~\cite{H_Nakamura01}.

The qubit MZ intereferometer differs in a number of ways from an
optical interferometer. Firstly, instead of a photon, the
interferometry is performed using the quantum state of a qubit.
Secondly, in the qubit, we have the interference of paths in phase
space rather than in coordinate space; the phase
$\Delta\theta_{12}$ (Eq.\ref{eq:phase}) is determined by the qubit
level splitting, which plays the role of the optical path length.
Finally, being more fragile than photons and easier to decohere,
qubit states can be manipulated in a coherent fashion only at
relatively short time scales.

We use a periodic driving signal, a harmonic variation of the
qubit detuning $\varepsilon(t)$:
\begin{equation}\label{eq:h(t)}
 {\cal H}=-\frac12 \left( \Delta\sigma^x +\epsilon(t)\sigma^z\right),\quad
 \epsilon(t)=\epsilon_0 + A_{\textrm{rf}}\cos\omega t,
\end{equation}
with $\Delta$ the tunnel splitting, $\epsilon_0$ the detuning
proportional to dc flux bias, and $A_{\textrm{rf}}$ the rf field
amplitude proportional to the rf flux bias (see supplementary
material).
%with $\Delta$ the energy splitting at the avoided crossing,
%$\epsilon_0$ the dc energy-detuning, and $A_{\textrm{rf}}$ the rf
%driving-field amplitude in units of energy.
In this case (Fig.\ref{fig:MZinterferometer}B) we have cascaded LZ
transitions which occur when the driving amplitude exceeds
detuning, giving rise to the inteference fringes at
$A_{\textrm{rf}}>|\epsilon_0|$ (Fig.\ref{fig:MZinterferometer}C).
While the phase $\Delta\theta_{12}$ equals the shaded area in
Fig.~\ref{fig:MZinterferometer}B and is dependent on
$A_{\textrm{rf}}$, the total phase gained over one period,
$\theta=\frac1{\hbar}\oint\epsilon(t)dt=2\pi\epsilon_0/\hbar\omega$,
equals the difference of the shaded and unshaded areas and is
independent of $A_{\textrm{rf}}$. As consecutive pairs of LZ
transitions (consecutive MZ interferometers) interfere
constructively when $\theta=2\pi n$, the fringes will appear
around the resonance detuning values
\begin{equation}\label{eq:resonance}
 \epsilon_{0,n}=n h\nu,\quad n=0,\,1,\,2,...,\quad \nu=\omega/2\pi .
\end{equation}
Another interpretation of this condition is that the sequential LZ
transitions excite multiphoton resonances.

While coherent multiphoton resonances between discrete states of
an rf-driven charge qubit have been
reported~\cite{NakamuraTsai99,Nakamura01}, and multiphoton
transitions used to drive Rabi oscillations in a flux
qubit~\cite{Saito05,Yu05}, in these works, as well as in the
earlier work on quantum dot
systems~\cite{Kouwenhoven94,Fujisawa97}, only a few-photon
transitions could be observed, with coherence quickly weakening at
increasing rf amplitude~\cite{Saito04a}. In contrast, we were able
to observe coherent resonances of very high order, up to $n=20$,
which requires driving the system at a high rf amplitude. The
fringes
%%LL displayed by these resonances
for high $n$ are as clear as those for $n\simeq 1$, indicating
that the qubit preserves a substantial amount of coherence even in
the strongly-driven regime.

We realize a tunable artificial atom with a niobium
persistent-current qubit (Fig.~\ref{fig:set_up}A), a
superconducting loop interrupted by three Josephson
junctions~\cite{Orlando99a}. When the qubit loop is threaded with
a magnetic flux $f_{\textrm{q}} \approx \Phi_0/2$, the system
exhibits a double-well potential-energy landscape
(Fig.~\ref{fig:qubit_potential_well}). The classical states of the
wells are persistent currents $I_{\textrm{q}}$ with opposing
circulation, described by energy bands $\pm \epsilon_0/2 = \pm
I_{\textrm{q}} \Phi_0 \delta f_{\textrm{q}}$ linear in the flux
detuning $\delta f_{\textrm{q}} \equiv f_{\textrm{q}} - \Phi_0/2
$. The double-well barrier allows quantum tunneling of strength
$\Delta$, opening the avoided crossing near $\delta f_{q} = 0$
(Fig.~\ref{fig:MZinterferometer}A). Detuning the flux tilts the
double well and, thereby, modifies its eigenenergies and
eigenstates. The qubit states are read out using a DC-SQUID, a
sensitive magnetometer that distinguishes the flux generated by
the circulating currents. The device was fabricated at MIT Lincoln
Laboratory (see supplementary material).

We drove transitions between the qubit states by applying a 1
$\mu$s rf pulse (Fig.~\ref{fig:set_up}B) at frequency $\nu$ and
rf-source voltage $V_{\textrm{rf}}$~\cite{voltage_explanation}.
% \propto A_{\textrm{rf}}$%
After a short ($\approx$10 ns) delay, we read out the qubit state
by driving the DC SQUID with a 20 ns ``sample'' current
$I_{\textrm{s}}$ followed by a 20 $\mu$s ``hold'' current. The
SQUID
will switch to its normal state voltage % and generate a voltage
$V_{\textrm{s}}$ if
 $I_{\textrm{s}} > I_{\textrm{sw,0}}$
 ($I_{\textrm{s}} > I_{\textrm{sw,1}}$),
%%LL the SQUID switching current corresponding to
%%LL qubit state $|0\rangle$ ($|1\rangle$).
corresponding to qubit state $|0\rangle$ and $|1\rangle$. By
sweeping the sample current and flux detuning while monitoring the
presence of a SQUID voltage over many trials, a cumulative
switching-distribution function is generated, revealing the
``qubit step'' (Fig.~\ref{fig:set_up}C). At specific values of
flux detuning, the rf field at $\nu = 1.2$ GHz becomes resonant
with the energy level separation, allowing $n$-photon absorption,
Eq.\ref{eq:resonance}; this results in a partial population
transfer between the qubits states, manifest as regularly spaced
``spikes'' in Fig.~\ref{fig:set_up}C. We obtain 1D scans of the
``switching probability'' $P_{\textrm{sw}}$ (the population of
state $|0\rangle$) shown in Fig.~\ref{fig:set_up}D by following a
flux-dependent sample current $I_{\textrm{sw,0}} < I_{\textrm{s}}
< I_{\textrm{sw,1}}$ (dash-dotted line in Fig.~\ref{fig:set_up}C).
Such 1D scans are then accumulated as a function of the rf source
parameters $V_{\textrm{rf}}$ (Fig.~\ref{fig:Bessel_staircase}) and
$\nu$ (Fig.~\ref{fig:freq_dependence}).

The switching probability $P_{\textrm{sw}}$ (color scale) vs.
qubit flux detuning $\delta f_{\textrm{q}}$ and voltage
$V_{\textrm{rf}}$ at frequency $\nu=1.2$ GHz is shown in
Fig.~\ref{fig:Bessel_staircase} (see supplementary material). The
$n$-photon resonances, labelled by $n=1 \ldots 20$, exhibit MZ
interference fringes (I$\ldots$VI) as a function of
$V_{\textrm{rf}}$. As we show below, the fringes exhibit a
Bessel-function dependence, $J_{n}(\lambda)$, so we call the
step-like pattern in Fig.~\ref{fig:Bessel_staircase} a ``Bessel
Staircase''. For each of the $n$-photon resonances, we took a
higher-resolution scan (e.g., Fig.~\ref{fig:Bessel_staircase}
inset) and fitted the resonance areas and widths in
Fig.~\ref{fig:Bessel_fits} (see supplementary material).

Multi-photon transitions at the resonances
(Eq.~\ref{eq:resonance}) in the strong driving regime,
$|A_{\textrm{rf}}|,h\nu\gg \Delta$, occur via fast LZ transitions.
The notion of quasistationary qubit levels
$\pm\frac12(\epsilon^2_0+\Delta^2)^{1/2}$ is inadequate in this
regime and, instead, we use a different approach, transforming the
Hamiltonian (Eq.~\ref{eq:h(t)}) to a nonuniformly rotating frame,
${\cal H}=e^{-(i/2)\phi(t)\sigma_z}{\cal
H}'e^{(i/2)\phi(t)\sigma_z}$, where $\phi(t)=\lambda\sin\omega t$
with dimensionless rf field amplitude
\begin{equation}\label{eq:lambda} \lambda=A_{\textrm{rf}}/h\nu \end{equation}
The rf field disappears from the detuning term, reappearing as a
phase factor of the off-diagonal term: $ {\cal H}'= -\frac12
\left( \varepsilon_0\sigma^z + \Delta e^{-i\phi(t)}\sigma^+ + {\rm
h.c.}\right) $. Given that $\Delta\ll h\nu$, near the $n$-th
resonance $nh\nu \simeq \epsilon_0$ we can replace the phase
factor $e^{-i\phi(t)}$ by its $n$-th Fourier harmonic,
$J_n(\lambda)e^{-in\omega t}$, with $J_n$ the Bessel function. The
resulting effective Hamiltonian is of a ``rotating field'' form:
\begin{equation}\label{eq:Hn(t)}
 {\cal H}'\approx {\cal H}_n =
    -\frac{1}{2}
   \left(
    \matrix{
        \epsilon_0 & e^{-in\omega t} \Delta_n \cr
         e^{in\omega t}\Delta_n^\ast & -\epsilon_0}
    \right)
\end{equation}
with $\Delta_n = \Delta J_n(\lambda)$. The resonance approximation
(Eq.~\ref{eq:Hn(t)}) describes transitions at an arbitrary ratio
$A_{\textrm{rf}}/h\nu$. Standard Rabi dynamics analysis of the
Hamiltonian (\ref{eq:Hn(t)}) with the initial state $|0\rangle$
gives the time-averaged occupation probability of the excited
state $P^{(n)}_{\textrm{sw}}=\frac12 |\Delta_n|^2/((\epsilon_0-n
h\nu)^2+|\Delta_n|^2)$. This expression predicts Lorentz-shaped
resonances of width $\delta \epsilon=|\Delta_n|$. The result, a
sum of independent contributions with different $n$,
\begin{equation}\label{eq:sum_of_peaks}
 P_{\textrm{sw}}=\frac12 \sum_{n}
    \frac{|\Delta_n|^2}{(\epsilon_0-n h\nu)^2+|\Delta_n|^2} ,
\end{equation}
is displayed in Fig.~\ref{fig:MZinterferometer}C. The agreement
with the observed resonances is striking: the oscillations in rf
power, described by $J_n(\lambda)$, accurately predict both the
overall profile of the fringes (Fig.~\ref{fig:Bessel_staircase})
and the fine details, such as positions of the nodes
(Fig.\ref{fig:Bessel_fits}).

In the frequency dependence of $P_{\textrm{sw}}$ for voltages
$V_{\textrm{rf}} = 71 \textrm{ mV}_{\textrm{rms}}$ and
$V_{\textrm{rf}} = 7.1 \textrm{ mV}_{\textrm{rms}}$
(Fig.~\ref{fig:freq_dependence}), the resonances approach the
qubit step as frequency decreases, in accordance with the linear
energy vs. flux-detuning dependence. MZ interference fringes are
again visible. The number of resonances increases at low
frequencies, due primarily to the frequency dependence of
$\lambda$ and, in lesser part, a frequency-dependent mutual
coupling.

Our analysis of peak profile accounts for the relaxation and
dephasing, as well as for the inhomogeneous broadening due to
low-frequency noise. These effects can be separated from one
another by considering the peak areas $A_n$ which, in contrast
with the widths of the resonances $w_n$, are not affected by
inhomogeneous broadening. The standard Bloch approach
yields~\cite{Abragam}
\begin{equation}
     \label{eq:areas}
 A_n = \frac{T_1 \Delta_n^2}{4\sqrt{T_1 T_2 \Delta_n^2+1}}
        ,\quad
 w_n = \frac{1}{\pi T_2^\ast}+\frac{\sqrt{T_1T_2\Delta_n^2+1}}{\pi T_2}
\end{equation}
with $T_{1,2}$ the longitudinal and transverse relaxation times,
and $T_2^\ast$ describing the inhomogeneous broadening. These are
aggregate relaxation times averaged over the periodic qubit
detuning, which, in the operating limit $\varepsilon_0 \gg
\Delta$, tends to overestimate $T_1$ and underestimate $T_2$
compared with their values at the avoided crossing.

Fig.~\ref{fig:Bessel_fits}A shows the Bessel dependence of the
$n=1$ and $n=5$ resonance areas fit by Eq.~\ref{eq:areas}
including (blue) and omitting (red) times  $T_{1,2}$. The
corresponding resonance widths and their fittings are shown in
Fig.~\ref{fig:Bessel_fits}B. Fig.~\ref{fig:Bessel_fits}C and
Fig.~\ref{fig:Bessel_fits}D show the resonance area and width
respectively for 10 resonances, including $n=19$. Fitting the
areas and widths yields self-consistent estimates: $T_1 \approx
20$ $\mu$s, $T_2 \approx 15 - 25$ ns, $T_2^* \approx 5-10$ ns, and
$\Delta / \hbar \approx (2 \pi)4$~MHz. The $T_1$, $T_2$, and
$\Delta$ estimates are similar to those reported in
Ref.~\cite{Yu05}. The nearly linear behavior at the nodes of $J_n$
[see Fig.~\ref{fig:Bessel_fits}A] indicates that the decoherence
is small compared with the splitting: $T_1 T_2 (\Delta/ \hbar)^2
\approx 250$ for $n=1$, and decreases slightly for $n=5$. The
fit/data discrepancy for the first fringe for $n=1$, which
disappears as $n$ increases, is traced to $\sim 20 \%$ thermal
population of the excited state due to its proximity to the qubit
step (see supplementary material).

This MZ interferometry technique finds application to qubit
characterization and model validation, two increasingly important
research areas in quantum information science. In addition to
coherence times, which can be obtained by multiple means, MZ
interferometry allows the direct calibration of the microwave
amplitude driving the qubit through the Bessel argument $\lambda$;
we found the rf mutual coupling to be $M_{\textrm{q}}=100 \pm
2\textrm{ fH} $ over all 20 resonances. The agreement between the
two-level Hamiltonian in Eq.~\ref{eq:h(t)} and the observed
resonances $n= 1 \ldots 10$ in Fig.~\ref{fig:Bessel_staircase} is
remarkable. The MZ technique also reveals shortcomings of the
two-level model at strong driving.
 For example, the influence of a second MZ
interferometer at the avoided crossing between the first and
second excited states results in the moir\'e-like pattern observed
for resonances $n>12$. We also note here an observed ($\sim 0.1$
GHz) shift in the resonance positions at strong driving
[Fig.~\ref{fig:Bessel_staircase} inset]. Both effects require the
presence of higher-excited states modelled by the full qubit
Hamiltonian~\cite{Orlando99a,Yu05}. The high stability and
coherence of the strongly driven qubit, even at $n=20$ photon
transitions, illustrates not only the potential for nonadiabatic
control methods~\cite{H_Nakamura01}, but also indicates the high
potential of niobium devices fabricated in a fully-planarized,
scalable process for superconductive quantum computation.

%We thank V. Bolkhovsky, G. Fitch, D. Landers, E. Macedo, R.
%Slattery, and T. Weir at MIT Lincoln Laboratory; and D. Berns, J
%Habif, and D. Nakada at the MIT campus for technical assistance.
%We thank D. Cory, S.~E. Harris, A.~J. Kerman, and S. Lloyd for
%helpful discussions.
%%KKB contributed to this work while on staff at MIT Lincoln Laboratory.
%This work was supported by AFOSR (F49620-01-1-0457) under the
%DURINT program. The work at Lincoln Laboratory was sponsored by
%the US DoD under Air Force Contract No. FA8721-05-C-0002.
%\bibliography{scibib}

\begin{thebibliography}{10}
\bibitem{voltage_explanation}
$V_{\textrm{rf}}$ is the rf voltage at the source generator. The
rf field amplitude $A_{\textrm{rf}}$ which drives the qubit is
parameterized in units of energy, and it is proportional to
$V_{\textrm{rf}}$. The proportionality constant can be inferred
from the intereference fringe spacing and the Bessel argument
$\lambda$.
\end{thebibliography}

\appendix
\section*{Supplementary Material}

\paragraph{Device fabrication and parameters}
The device was fabricated at MIT Lincoln Laboratory on 150 mm
wafers in a fully-planarized niobium trilayer process with
critical current density $J_{\textrm{c}} \approx 160$ $\textrm{A}/
\textrm{cm}^2$. The qubit's characteristic Josephson and charging
energies are $E_{\textrm{J}} \approx (2\pi\hbar)300$ GHz and
$E_{\textrm{C}} \approx (2\pi\hbar)0.65$ GHz respectively, the
ratio of the qubit JJ areas is $\alpha \approx 0.84$, and the
tunnel coupling $\Delta\approx (2\pi\hbar)0.005$ GHz. The qubit
loop area is 16 $\times$ 16 $\mu$m$^2$, and its self inductance is
$L_{\textrm{q}} \approx 30$ pH. The SQUID JJs each have critical
current $I_{\textrm{c0}} \approx 2$ $\mu$A. The SQUID loop area is
20 $\times$ 20 $\mu$m$^2$, and its self inductance is
$L_{\textrm{SQ}} \approx 30$ pH. The mutual coupling between the
qubit and the SQUID is $M \approx 25$ pH. An experimental
spectroscopic analysis of the first five energy levels of this
qubit can be found in Ref.~\cite{Yu05}.

\paragraph{Implementation}
The experiments were performed in a dilution refrigerator with a
12 mK base temperature. The device was magnetically shielded with
4 Cryoperm-10 cylinders and a superconducting enclosure. All
electrical leads were attenuated and/or filtered to minimize
noise.

The data were obtained at a repetition rate of 200 Hz, allowing
approximately 5 ms for qubit equilibration (initialization)
between trials. In Fig.\ref{fig:Bessel_staircase}, 3000 steps were
taken along the flux axis and 250 steps were taken along the
amplitude axis. Each point comprised an average of 100 trials. The
data set was accumulated continuously over approximately 4.5 days.
The Fig.~\ref{fig:Bessel_fits} data set was obtained from 2D scans
acquired at the same step size in flux and amplitude, but over
extended amplitude ranges for higher-$n$ transitions and with each
point averaging 2000 trials. Although the onset of the 20-photon
resonance is difficult to observe in
Fig.~\ref{fig:Bessel_staircase} due to the color shading, we
clearly observed it when scanning to higher amplitudes. The
plotting and fitting algorithms used to obtain
Fig.~\ref{fig:Bessel_staircase} and Fig.~\ref{fig:Bessel_fits}
were fully scripted and identically applied to each resonance.

%\section*{Supporting Text}

\paragraph{Temperature effects}
The thermal energy in frequency units is $k_{\textrm{B}} T / h
\approx 1$ GHz as determined from the qubit step in Fig.~2D. The
qubit is dc-biased, prepared, and measured far from the avoided
crossing, $\varepsilon_{0,n} \sim \hbar \nu_n
> k_{\textrm{B}} T \gg \Delta$. The qubit can be operated in this limit,
because the equilibrium population is predominantly in the ground
state (particularly for $n>3$), and the coherence time scales are
much longer than the rf drive oscillation period over which the
Mach-Zehnder physics occurs. Where residual excited state
population does exist, e.g., $n=1$ due to its proximity to the
qubit step, it acts to offset the Bessel function dependence of
the spectroscopy peaks. Since population inversion should not
occur for our qubit in the limit of steady-state spectroscopy, the
first lobe height is clamped at $\sim 50 \%$  (in our experiment,
the other lobes do not exceed $50 \%$ even with the offset). When
the offset is removed and the Bessel fit is made, the data/fit
discrepancy is observable for the first lobe of $n=1$. The effect
is less pronounced for $n=2$ and $n=3$ as they are further from
the qubit step with a smaller and smaller equilibrium
excited-state population.

\paragraph{Differentiation from previous work}
The multiphoton transitions observed here are distinct from
Tien-Gordon photon-assisted transport~\cite{Tien63a} or
above-threshold ionization of atoms~\cite{Paulus01a}, both of
which utilize a continuum of final states. Furthermore, the Bessel
dependence is a quantum-coherent interferometric effect and,
thereby, differs from the classical, non-linear phase-locking
mechanism that may also lead to a Bessel dependence, e.g., as with
Shapiro steps~\cite{Shapiro68a}. The Landau-Zener and Mach-Zehnder
physics observed here is a coherent interferometric effect and,
thus, differs from previous Landau-Zener work in the incoherent
limit~\cite{Ilichev04a}.

In our previous work in Ref.~\cite{Yu05}, we observed
single-photon, two-photon, and three-photon transitions by weakly
driving the qubit at frequencies $\nu=1 \ldots 20$ GHz. In the
present work, we are able to observe up to $n=20$ photon
transitions by strongly driving the qubit at relatively low
frequencies $\nu=0.2 \ldots 1.4$ GHz, allowing sufficient
``headroom'' for multi-photon transitions to high-order between
the ground and first-excited states.

\bibliographystyle{Science}

% Following is a new environment, {scilastnote}, that's defined in the
% preamble and that allows authors to add a reference at the end of the
% list that's not signaled in the text; such references are used in
% *Science* for acknowledgments of funding, help, etc.

\begin{scilastnote}
\item We thank V. Bolkhovsky, G. Fitch, D. Landers, E. Macedo, R.
Slattery, and T. Weir at MIT Lincoln Laboratory; and D. Berns, J
Habif, and D. Nakada at the MIT campus for technical assistance.
We thank D. Cory, S.~E. Harris, A.~J. Kerman, and S. Lloyd for
helpful discussions.
%KKB contributed to this work while on staff at MIT Lincoln Laboratory.
This work was supported by AFOSR
(F49620-01-1-0457) under the DURINT program. The work at Lincoln
Laboratory was sponsored by the US DoD under Air Force Contract
No. FA8721-05-C-0002.
\end{scilastnote}

\clearpage
\begin{figure}[t]
% \caption{}
% \epsfig{file=fig1_oliver2005_v7.eps,width=6.5in}
 \caption{
 Mach-Zehnder interference in a strongly driven qubit.
 (\textbf{A}) Starting at the dot marker, the qubit state is swept by
    an rf field.
    After a LZ transition at the first avoided crossing (time $t_1$),
    the resulting superposition state of $|0\rangle$ and $|1\rangle$ (dashed lines) accumulates
    a phase $\Delta\theta_{12}$ (shaded region), and interferes at the
    return LZ transition (time $t_2$).
    The qubit state is subsequently driven away from the avoided crossing and
    then returns to the starting flux position. This single period of qubit evolution is
    a single MZ interferometer. Depending on the interference phase $\Delta \Theta_{12}$,
    amplitude may build in the excited state.
 (\textbf{B})  The corresponding qubit energy variation induced by a periodic rf field,
    Eq.\ref{eq:h(t)}, results in an equivalent optical cascade of MZ
    interferometers (MZ\#1-\#3, top) with resonance condition Eq.~\ref{eq:resonance}.
 (\textbf{C}) The population of the qubit excited state,
    Eq.\ref{eq:sum_of_peaks}, as a function of rf amplitude
    $A_{\textrm{rf}}$ and detuning $\epsilon_0$. Note the interference
    fringes (I, II, III, ...) at $A_{\textrm{rf}} > \epsilon_0$ and
    the multiphoton resonances at $\epsilon_0=n h \nu$.
    }  \label{fig:MZinterferometer}
\end{figure}

%\clearpage
\begin{figure}[t]
% \caption{}
% \epsfig{file=fig2_oliver2005_v2.eps}%,width=3.5in}
 \caption{
Multiple resonances in a strongly-driven flux qubit.
 (\textbf{A})  Circuit schematic of the three-junction flux qubit (inner loop)
    with circulating current $I_{\textrm{q}}$ and the
    DC SQUID readout (outer loop); Josephson junctions are indicated with an
    $\times$. A time-dependent flux $f(t) \propto \varepsilon(t)$ threading the qubit is a
    sum of the flux bias due to the dc current $I_b$ and a pulsed ac
    current at frequency $\nu$ irradiating the qubit and driving
    transitions between its quantum states.
    The SQUID is shunted by two 1\,pF
    capacitors to lower its resonance frequency. Resistors
    mark the environmental impedance isolating the SQUID.
 (\textbf{B})  The time sequence for the rf pulse
    (duration 1 $\mu$s and rf-source voltage $V_{\textrm{rf}}$)
    and SQUID sample current $I_{\textrm{s}}$.
    A repetition period of 5 ms allows for equilibration between trials.
 (\textbf{C})  A cumulative switching-probability distribution of the qubit as a
    function of $I_{\textrm{s}}$ and the qubit flux detuning $\delta f_{\textrm{q}}$
    under
    rf excitation at $V_{\textrm{rf}} \approx 0.12 \textrm{ V}_{\textrm{rms}}$
    and $\nu = 1.2$ GHz. Multi-photon transitions are observed between the qubit
    states $|0\rangle$ and $|1\rangle$ and are symmetric about the
    qubit step
    ($\delta f_{\textrm{q}}=0$ $\textrm{m}\Phi_0$).
 (\textbf{D})  The 1D switching probability $P_{\textrm{sw}}$
    extracted from Fig.~\ref{fig:set_up}C (dash-dotted line scan).
    }  \label{fig:set_up}
\end{figure}

%\clearpage
\begin{figure}[t]
% \caption{}
% \epsfig{file=fig3_oliver2005_v3_alt_eps.eps} %,width=3.5in}
 \caption{
Multi-photon interference fringes: a ``Bessel-staircase''.
 Switching probability $P_{\textrm{sw}}$ is plotted as a function of qubit flux
 detuning $\delta f_{\textrm{q}}$ and
 %WDO rf-source
 voltage $V_{\textrm{rf}}$ at frequency
 $\nu = 1.2$ GHz.
 %WDO amplitude
 $n$-photon resonances are labelled $1 \ldots 20$. Each $n$-photon resonance
 exhibits oscillations in $P_{\textrm{sw}}$ due to a MZ-type
 quantum interference that results in a Bessel
 dependence $J_{n}(\lambda)$ with $\lambda$ the rf amplitude
 scaled by $h\nu$, Eq.(\ref{eq:lambda}).
 Roman numerals mark the interference fringes of $J_{n}(\lambda)$ (solid
 white lines). The $n$-photon resonances are symmetric about the qubit step ($0 \textrm{ m}\Phi_0$).
 Inset: close-up of the $n=4$
 photon resonance.
 }  \label{fig:Bessel_staircase}
\end{figure}

\begin{figure}[t]
 \caption{Frequency dependence of the multiphoton interference fringes.
  Qubit switching probability $P_{\textrm{sw}}$ is plotted as a function of
  frequency $\nu$
  and flux detuning $\delta f_{\textrm{q}}$ in the limit of
 (\textbf{A}) strong driving ($V_{\textrm{rf}} = 71 \textrm{ mV}_{\textrm{rms}}$) and
 (\textbf{B}) weak driving ($V_{\textrm{rf}}= 7.1 \textrm{ mV}_{\textrm{rms}}$). The $n$-photon resonances approach the qubit step as
  frequency decreases. Symmetric patterns in the $J_{n}(\lambda)$ peaks and
  valleys due to the MZ-type quantum interference are clearly observed.
} \label{fig:freq_dependence}
\end{figure}

%\clearpage
\begin{figure}[t]
% \caption{}
% \epsfig{file=fig5_oliver2005_v2.eps,width=6in}
 \caption{
Analysis of the resonance area and width.
 (\textbf{A}) Resonance area $A_n$ versus %WDO rf-source
    voltage $V_{\textrm{rf}}$ for the
    $n=1$ and $n=5$ photon transitions. The Bessel-dependence
    $J_{n}(\lambda)$ is observed over several lobes. The data are best
    fit by functions that include decoherence (blue line) rather than
    omit it (red line). As illustrated by the insets, decoherence
    becomes more pronounced as photon number increases.
 (\textbf{B}) The resonance width $w_n$ versus %WDO rf-source
    voltage $V_{\textrm{rf}}$ for the $n=1$ and $n=5$
    also exhibits a Bessel dependence.
 The area (\textbf{C}) and the width (\textbf{D}) plotted for resonances $n=1 \ldots 9$ and
 $n=19$.
 } \label{fig:Bessel_fits}
\end{figure}

\begin{figure}[t]
 \caption {Qubit double-well potential and the corresponding energy band diagram for the two-lowest qubit
 states. The circulating currents of the qubit correspond to the states $|0\rangle$ and $|1\rangle$ of
 the uncoupled left and right wells. These single-well states are coupled with tunneling strength $\Delta$
 by the potential barrier; the eigenstates of the coupled system are denoted ``e'' and ``g''. Varying the flux detuning $\delta
 f_q$ tilts the double-well potential and, thereby, modifies the
 energy band structure of the qubit. Far from $\delta
 f_q = 0$, the eigenstates and eigenenergies resemble the
 single-well states and energies. Near $\delta
 f_q = 0$, the system exhibits an avoided
 crossing.
}
 \label{fig:qubit_potential_well}
\end{figure}

%\clearpage

\end{document}